\documentstyle[11pt, amssymb]{article}
  \newcommand {\nc}{\newcommand}
  \nc{\eq}{\begin{equation}}
  \nc{\en}{\end{equation}}
  \nc{\eqa}{\begin{eqnarray}}
  \nc{\ena}{\end{eqnarray}}
  \newtheorem{definition}{Definition}

  \def\ot{\otimes}
  \def\nat{{\mathbb N}}
  \def\intg{{\mathbb Z}}
  \def\real{{\mathbb R}}
  \def\complex{{\mathbb C}}
  \def\Alg{{\mathcal A}}
  \def\Hs{{\mathcal H}}
  \def\Drc{{\mathcal D}}

  \nc {\norm}[1]{||{#1}||}
 \title{Geometric Origin of Staggered Fermion: Direct Product $K$-Cycle}
 \author{Jian Dai
  \thanks{E-mail: jdai@mail.phy.pku.edu; Postal address: Room 2082, Building 48, Peking University, Beijing, P. R. China, 100871},
  Xing-Chang Song
  \thanks{E-mail: songxc@ibm320h.phy.pku.edu.cn; Postal address: Theory Group, Department of
  Physics, Peking University, Beijing, P. R. China, 100871}\\
  Theory Group, Department of Physics, Peking University
  }
 \date{May 14th, 2001}
 
\begin{document}
  \noindent
  \maketitle
  \begin{abstract}
   Staggered formalism of lattice fermion can be cast into a form of direct product $K$-cycle
   in noncommutative geometry. The correspondence between this staggered $K$-cycle and a canonically defined $K$-cycle
   for finitely generated abelian group where lattice appears as a special case is proved.
  \end{abstract}
  \section{Introduction and Preliminary}
   Staggered formalism for lattice fermion is one of the earliest
   solutions to the puzzle of species doubling in lattice field
   theory (LFT) \cite{St1}; thorough exploration of this formalism
   were carried out in a series of work, especially on the problems of
   flavor interpretation and gauge coupling \cite{St2}; the
   dynamical properties of staggered fermion were considered in \cite{St3}. Recently, a
   nontrivial correspondence between staggered Dirac operator and
   noncommutative geometry (NCG) was figured out \cite{DS1};
   however, this correspondence was still a conjecture in general
   since rigid proof was just given for lattice whose dimension is one, two
   or four. In this letter, with the power of NCG being fully
   employed,
   a proof of this correspondence for any dimensional lattice is
   presented. This article is organized in the following way. A concise
   introduction of the central objects for NCG, {\it $K$-cycles}, and their direct product will be given below.
   In Sect.\ref{a},
   direct product $K$-cycle for finitely generated abelian group
   is introduced, with lattice being treated as a
   special example. In Sect.\ref{b}, staggered formalism is also cast in the
   shape of direct product $K$-cycle, so the proof of the above-mentioned correspondence
   is reduced to show that the correspondence exists within each factor. Some remarks and
   discussions are put in Sect.\ref{c}.\\

   Comprehensive introduction to NCG can be found in \cite{NCG1}.
   Only concepts relevant to our work are recalled below.
   \begin{definition}
    An even $K$-cycle in Connes' operator-algebraic approach towards NCG is
    presented as a quadruple $(\Alg, \Hs, \Drc, \gamma)$, in which $\Alg$ is a
    pre-$C^\ast$ algebra being represented faithfully and unitarily
    on a separable Hilbert space $\Hs$ by $\pi$, Dirac operator $\Drc$ is a selfadjoint
    operator on $\Hs$ with compact resolvent, and $\gamma$ is a selfadjoint involution on $\Hs$, providing
    $\Hs$ with a $\intg_2$-grading such that
    $[\pi(\Alg),\gamma]=0$, $\{\Drc, \gamma\}=0$.
   \end{definition}
   A collection of axioms is imposed on every $K$-cycle, being even or not, such
   that when $\Alg$ is commutative, a $K$-cycle will recover a
   spin-manifold \cite{Co1}.
   \begin{definition}\label{Def2}
    Direct product of two even $K$-cycle $(\Alg_i, \Hs_i, \Drc_i, \gamma_i),
    i=1,2$ is another even $K$-cycle $(\Alg, \Hs, \Drc, \gamma)$ where
    $\Alg=\Alg_1\ot\Alg_2$, $\Hs=\Hs_1\ot\Hs_2$,
    $\gamma=\gamma_1\ot\gamma_2$,
    $\Drc=\Drc_1\ot Id+\gamma_1\ot\Drc_2$.
   \end{definition}
   In a rigid sense,
   axiomatic $K$-cycle does not exist for finite sets or lattices \cite{LattNCG1},
   unless $\Hs$ is large enough, though being unnatural.
   Therefore, the term $K$-cycle in this paper is in a weak form.
  \section{Direct Product $K$-Cycle over Finitely Generated Abelian Group}\label{a}
   Lattice can be fitted into a more general category, the category of finitely
   generated abelian group (FGAG). Since the classification of FGAG is
   totally clear in group theory, it is possible to assign canonical
   $K$-cycle over this category. In fact, any FGAG can be uniquely
   decomposed as a direct product group whose factors are either $\intg$
   or $\intg_{p^k}$, where $p$ is a prime number and
   $k=1,2,3,\ldots$ \cite{Grp1}. Hence, if a canonical $K$-cycle is settled
   for each fundamental block, $\intg$ and $\intg_{p^k}$, then $K$-cycle for any FGAG
   can be naturally defined by direct product of $K$-cycles of each factor
   group. Note importantly that $\intg_2$ factors will not be considered in this letter, due to
   their speciality; treatment of differential structure on direct product of $\intg_2$ can be found in
   \cite{Mj}.\\

   Let $G$ be a FGAG without $\intg_2$ factors,
   \[
    G=\bigotimes_{f=1}^N G_f, G_f\in\{\intg, \intg_{p^k}:k\in\nat, \mbox{if }p>2; k\in\nat\setminus\{1\}, \mbox{if }p=2\}
   \]
   Fixing one $f$, a canonical $K$-cycle is specified as below.
   $\Alg_f$ is the algebra of complex functions over
   $G_f$; if $G_f$ is $\intg$, then the functions are required to be bounded. $\Hs_f=l^2(G_f)\ot\complex^2$ gives the Hilbert
   space; the choice of factor $\complex^2$ will be clarified
   below.
   $\Alg_f$ is represented on $\Hs_f$ by multiplication $\pi_f(u)=u\ot Id$, $\forall
   u\in\Alg_f$. The generator of $G_f$ is write as
   $T_f$, and it induces the regular representation of $G_f$ on $\Alg_f$
   in which the image of $T_f$ is denoted as $R_{T_f}$. Define formal partial
   derivatives $\partial_{T_f}=R_{T_f}-Id$, and
   $\partial_{T_f^{-1}}=R_{T_f^{-1}}-Id$, then Dirac operator is
   specified as
   \eq\label{Dof}
    \Drc_f=b^\dag\partial_{T_f}+b\partial_{T_f^{-1}}
   \en
   where $b,b^\dag$ is a pair of standard fermionic annihilation/creation operators, being represented irreducibly on $\complex^2$.
   Grading $\gamma_f$
   is $[b^\dag, b]/2$. One can verify that $\Drc_f^\dag=\Drc_f$, $\Drc^2_f=\partial_{T_f}\partial_{T_f^{-1}}$,
   $[\pi(\Alg_f), \gamma_f]=0$, and $\{\Drc_f, \gamma_f\}=0$.
   The direct product $K$-cycle on $G$ can be established straightforwardly. Namely following
   Definition \ref{Def2}, $\Alg[G]=\bigotimes_f \Alg_f$, $\Hs=\bigotimes_f\Hs_f$,
   \eq\label{Do}
    \Drc=\Drc_1\ot Id\ot Id\ot\ldots\ot Id+\gamma_1\ot\Drc_2\ot
    Id\ot\ldots\ot Id+\ldots +
    \gamma_1\ot\gamma_2\ot\ldots\ot\gamma_{N-1}\ot\Drc_N,
   \en
   \eq\label{Gd}
    \gamma=\gamma_1\ot\gamma_2\ot\ldots\ot\gamma_3.
   \en
   A d-dimensional lattice can be considered as $\intg^d$, so that
   the above canonical $K$-cycle for FGAG can be applied to this
   lattice simply.
  \section{Staggered Formalism as Direct Product $K$-Cycle}\label{b}
   Staggered Dirac operator has standard expression
   \eq\label{Dos}
    \Drc_S=i\eta^\mu\nabla_\mu
   \en
   where $\nabla_\mu$ is symmetric difference operator defined as
   $(T_\mu-T_{-\mu})/2$, $\eta^\mu$ is called {\it staggered
   phase} expressed as $(-)^{\sum_{\nu <\mu}x_\nu}$, and an additional ``$i$'' is inserted in this definition
   since here a selfadjoint convention is adopted instead of an
   anti-selfadjoint one. The main advantage of staggered formalism is that
   there exists a chirality $\gamma_S=(-)^{x_1+x_2+\ldots +x_d}$,
   such that $\{\Drc_S, \gamma_S\}=0$. Observe
   Eqs.(\ref{Do})(\ref{Gd}) and Eq.(\ref{Dos}), staggered formalism can be cast into
   a direct product $K$-cycle in the following way. Let
   $\intg^d=\bigotimes_{\mu=1}^d\intg_{[\mu]}$, and for each
   factor, let $\Alg_{[\mu]}$ be algebra of bounded functions on
   $\intg_{[\mu]}$, $\Hs_{[\mu]}$ be the Hilbert space which is
   the restriction of $\Alg_{[\mu]}$ by $l^2$-condition; Dirac operator
   for $\intg_{[\mu]}$ is just $i\nabla_\mu$ and grading is taken
   to be chirality along $\mu$-direction,
   $\gamma_{[\mu]}=(-)^{x_\mu}$. Then it is easy to check the direct
   product property of staggered formalism.
   $\Alg[\intg^d]=\bigotimes_\mu\Alg_{[\mu]}$,
   $\Hs_S=\bigotimes_\mu\Hs_{[\mu]}=\Alg[\intg^d]_{l^2}$, $\gamma=\gamma_S=\Pi^\ot_\mu\gamma_{[\mu]}$,
   and very similar to Eq.(\ref{Do})
   \[
    -i\Drc_S=\nabla_1+\gamma_{[1]}\nabla_2+\gamma_{[1]}\gamma_{[2]}\nabla_3+\ldots
    +\gamma_{[1]}\gamma_{[2]}\ldots\gamma_{[d-1]}\nabla_d
   \]
   The canonical $K$-cycle $(\Alg[\intg^d],\Hs, \Drc,\gamma)$ of d-dimensional lattice defined in the last section is
   equivalent to the direction product $K$-cycle $(\Alg[\intg^d], \Hs_S, \Drc_S, \gamma_S)$ for staggered fermion,
   if and only if the
   equivalence holds for each factor, namely Dirac
   operator defined in Eq.(\ref{Dof}) is equivalent to $i\nabla$ for $\intg$.
   However, this statement has been shown in \cite{DS1} for one-dimensional lattice.
   For the integrality of this letter, this proof is rewritten
   below.\\

   Staggered $K$-cycle of one-dimensional lattice is
   $(\Alg[\intg], \Alg[\intg]_{l^2}, i\nabla, (-)^x)$. The
   spectral of $\Alg[\intg]$ correspondences to $\intg$ by
   Gelfand-Na\v{\i}mark theorem \cite{Pr}, namely there is
   bijection between pure states $\{|x\rangle \}$ over $\Alg[\intg]$ and $\intg$.
   Moreover, $\{|x\rangle\}$ also provides the Hilbert space with
   a basis. Now define fermionic operators $c^\dag$ by $c^\dag
   |2n\rangle =\sqrt{2}|2n+1\rangle$, $c^\dag|2n+1\rangle =0$, and
   $c$ by conjugation. It is obvious that $c^2=0$, $c^{\dag 2}=0$,
   $\{c,c^\dag\}=2$. Reparametrize $\{|x\rangle\}$ as
   $\{|n\rangle\rangle=|2n\rangle,
   c^\dag|n\rangle\rangle=|2n+1\rangle\}$, which is the reformulation of well-known ``half-spacing transformation''
   in \cite{St2}. Then one can check that
   under this parametrization,
   \eq\label{Doc}
    i\nabla={i\over 2}(c^\dag\partial_T^\dag-c\partial_T)
   \en
   where $\partial_T|n\rangle\rangle=|n+1\rangle\rangle -|n\rangle\rangle$, and
   $[c,\partial_T]=0$. So, identify that $c=ib^\dag$ in
   Eqs.(\ref{Dof})(\ref{Doc}), the correspondence between staggered fermion and
   NCG is proved for lattice of any dimension $d$.
  \section{Geometric Interpretation}\label{c}
   1) The result in the last section is not very striking in the sense that a lot of clues have been shown in
   in \cite{St2} where the so-called {\it spin
   diagonalization} tech were adopted extensively. What appears as a surprise is
   such a simplicity of the proof in the language of direct product $K$-cycle which is almost
   a trivial object in NCG.\\

   2) The proof within one factor inspires a geometric
   interpretation of staggered formalism. In fact, when a
   continuum space, say $\real^4$, is discretized, the tangent
   space at each point of the resulting lattice can be discretized also to be a
   linear space over character-$2$ field, namely for each
   direction, there are only two points in the tangent space,
   corresponding to two states $|n\rangle\rangle, c^\dag |n\rangle\rangle$, for all $n\in\intg$. The chirality
   can be expressed in operator form $\gamma=e^{i\pi c^\dag c}$ which is just {\it R-parity}
   in supersymmetry. \\

   3) $\gamma$ is essentially a distinguish between even and odd
   numbers, hence only exists globally for prime factor groups $\intg$ and
   $\intg_{2^k}, k=1,2,...$.\\

   {\bf Acknowledgements}\\
    This work was supported by Climb-Up (Pan Deng) Project of
    Department of Science and Technology in China, Chinese
    National Science Foundation and Doctoral Programme Foundation
    of Institution of Higher Education in China. J.D. is grateful
    for Dr. B-S. Wang on the help of group theory.

  
 \end{document}